\PassOptionsToPackage{prologue,dvipsnames}{xcolor}
\documentclass[sigconf]{acmart}

\usepackage[ruled,linesnumbered]{algorithm2e}
\usepackage{listings} 
\usepackage{bbding}
\usepackage{enumitem}
\usepackage{pifont}
\usepackage{multicol}
\newcommand{\appname}{{\sc Vesta}\xspace}
\newcommand{\baselinename}{{\sc Transfer}\xspace}

\AtBeginDocument{%
  \providecommand\BibTeX{{%
    \normalfont B\kern-0.5em{\scshape i\kern-0.25em b}\kern-0.8em\TeX}}}

\setcopyright{acmcopyright}
\copyrightyear{2023}
\acmYear{2023}
\acmDOI{XXXXXXX.XXXXXXX}

\acmConference[ICSE 2024]{46th International Conference on Software Engineering}{April 2024}{Lisbon, Portugal}

\begin{document}

\title[Exploiting Library Vulnerability via
 Migration Based Automating Test Generation]{Exploiting Library Vulnerability via \\ Migration Based Automating Test Generation
 }

\author{Zirui Chen}
\affiliation{%
  \institution{Zhejiang University}
  \country{China}
}
\email{chenzirui@zju.edu.cn}

\author{Xing Hu}
\authornote{Corresponding Author}
\affiliation{
  \institution{Zhejiang University}
  \country{China}
  }
\email{xinghu@zju.edu.cn}

\author{Xin Xia}
\affiliation{%
  \institution{Huawei}
  \country{China}
}
\email{xin.xia@acm.org}

\author{Yi Gao}
\affiliation{%
  \institution{Zhejiang University}
  \country{China}
}
\email{gaoyi01@zju.edu.cn}

\author{Tongtong Xu}
\affiliation{%
  \institution{Huawei}
  \country{China}
}
\email{xutongtong9@huawei.com}

\author{David Lo}
\affiliation{%
  \institution{Singapore Management University}
  \country{Singapore}
}
\email{davidlo@smu.edu.sg}

\author{Xiaohu Yang}
\affiliation{%
  \institution{Zhejiang University}
  \country{China}
}
\email{yangxh@zju.edu.cn}

\begin{abstract}
  In software development, developers extensively utilize third-party libraries to avoid implementing existing functionalities. When a new third-party library vulnerability is disclosed, project maintainers need to determine whether their projects are affected by the vulnerability, which requires developers to invest substantial effort in assessment. However, existing tools face a series of issues: static analysis tools produce false alarms, dynamic analysis tools require existing tests and test generation tools have low success rates when facing complex vulnerabilities.
  
  Vulnerability exploits, as code snippets provided for reproducing vulnerabilities after disclosure, contain a wealth of vulnerability-related information. This study proposes a new method based on vulnerability exploits, called \appname (Vulnerability Exploit-based
  Software Testing Auto-Generator), which provides vulnerability
  exploit tests as the basis for developers to decide whether to update
  dependencies. \appname extends the search-based test generation methods by adding a migration step, ensuring the similarity between the generated test and the vulnerability exploit, which increases the likelihood of detecting potential library vulnerabilities in a project.
  
  We perform experiments on 30 vulnerabilities disclosed in the past five years, involving 60 vulnerability-project pairs, and compare the experimental results with the baseline method, \baselinename. The success rate of \appname is 71.7\% which is a 53.4\% improvement over \baselinename in the effectiveness of verifying exploitable vulnerabilities. 
\end{abstract}


%   <concept>
%     <concept_id>10011007.10011006.10011072</concept_id>
%     <concept_desc>Software and its engineering~Software libraries and repositories</concept_desc>
%     <concept_significance>500</concept_significance>
%   </concept>
%   <concept>
%     <concept_id>10011007.10011074.10011784</concept_id>
%     <concept_desc>Software and its engineering~Search-based software engineering</concept_desc>
%     <concept_significance>500</concept_significance>
%   </concept>

% \ccsdesc[500]{Software and its engineering~Software libraries and repositories}
% \ccsdesc[500]{Software and its engineering~Search-based software engineering}


\maketitle

\section{Introduction}
%背景
Open-source libraries are widely used during software development ~\cite{Wang1, Kula1}. It is estimated that 96\% of software projects contain open source resources ~\cite{Synopsys1}. The widespread use of open-source libraries allows developers to reuse common functionalities, saving time and resources ~\cite{Yuan1, Kula1, Haryono1}. Similar to other software projects, open-source libraries may contain flaws ~\cite{Bavota1, Chen1}, which increase the possibility of attacks on software systems ~\cite{Chensen1, Tang1}. Vulnerabilities in open-source libraries can be particularly serious for the following reasons: 1) The vulnerabilities may propagate to dependent packages ~\cite{Decan1, Mir1}; 2) Vulnerabilities in open-source libraries can expose client applications to abuse ~\cite{Iannone1}. For example, in 2021, a remote code execution bug was disclosed in the log4j2 library, which affected millions of devices and required developers to upgrade dependency quickly ~\cite{Hong1, James1}.

%提出问题 因为开发者原因，很多项目的第三方漏洞都没有检测
When a new fix for existing bugs in a library is released, developers must decide if the dependency version should be updated. However, updating the dependency version may introduce conflicts into software ecosystems ~\cite{Bogart1}. For instance, the library API may break when fixing bugs, refactoring code, or adding features ~\cite{Wang1,kim1}, making it impossible for developers to upgrade the dependency version immediately. Due to this reason, a significant amount of projects are exposed to the library vulnerabilities ~\cite{Dallmeier1, Mirhosseini1, Wang1, Zhan1, Johannes1}.

%提出背景，很多第三方漏洞其实是无法被触发的
The inclusion of a vulnerable dependency in a project does not necessarily mean that the project is affected by the vulnerability ~\cite{Yuan1}. According to Zapata's research ~\cite{Elizalde1}, 73.3\% of projects that depend on vulnerable dependencies are secure. For a library vulnerability to be exploitable, it must satisfy two conditions: 1) the project must contain a control flow that calls the vulnerable function, and 2) the client project should be able to pass a crafted input that triggers the vulnerability ~\cite{Ali1}. For this reason, developers should check whether the project is affected by the vulnerability before updating the dependency version. This task can be particularly time-consuming in large projects ~\cite{Cobleigh1}.

%提出解决思路，已有的检测第三方漏洞的工具
To release developers from checking the vulnerabilities' exploitability in projects, existing tools are developed to detect whether the project is affected by the vulnerability. Dependency-based methods analyze vulnerable dependency versions to check potential vulnerabilities ~\cite{Elizalde1, Alfadel1, Decan1}. Call graph-based methods analyze the control flow of the project, checking if it contains a function call to the vulnerable function ~\cite{Ponta1, Nielsen1}. Dynamic detecting methods executing existing tests or generated tests, checking whether a control flow could reach the vulnerable function to detect the exploitable vulnerabilities ~\cite{Darius1, Ponta2}.

%提出问题，上述工具存在的问题，与基于测试生成的工作
However, the aforementioned methods will cause false alarms ~\cite{Cadariu1} as the lack of measuring whether client projects could construct inputs to trigger the vulnerabilities. To solve the problem, some researchers ~\cite{Hong1, Iannone1} focused on generating a test case to call the project APIs as attackers. The test could be an exploit if it could trigger the library vulnerability in the project. Recently, \baselinename ~\cite{Hong1} utilized vulnerability witness tests to overcome the lack of domain knowledge and the intrinsic complexity of exploiting vulnerabilities. In \baselinename, vulnerability witness tests are used to collect the trigger condition. By carving the libraries' test, \baselinename obtained the program state associated with the triggering of the vulnerability, which is added to the fitness function to evaluate the possibility of exploiting the library vulnerability.

%提出我们的思路
Similarly, we use vulnerability exploit code to collect domain knowledge for triggering. We collect a total of 747 projects that rely on the Jackson-Databind library, which is used to read content encoded in JSON or other data formats as well, as long as the parser and generator implementations exist ~\cite{Jackson1}. Jackson-Databind is associated with CVE-2019-14540 and an additional 46 vulnerabilities. We specifically chose 247 projects that include a call graph for the vulnerable function called \texttt{readValue}. Through manual examination of these call graphs, we discover that certain projects either passed parameters to the vulnerable function without making any alterations or made simple modifications, such as changing the parameter type or adding a substring. Based on the aforementioned conclusion, we propose a hypothesis that transferring parameter values extracted from vulnerability exploit to a specific call graph enables the vulnerable functions in the project to receive a parameter value capable of triggering the vulnerability.

%讲述方法
In this paper, we propose \appname to assess the exploitability of vulnerabilities by generating test cases that trigger the vulnerabilities. Instead of executing the witness test in \baselinename, we collect parameters from the exploit to ensure the trigger of vulnerabilities. For each vulnerability, we collect an exploit code snippet that provides domain knowledge for triggering the vulnerability. \appname extracts parameters from the exploit code and utilizes rules to modify it. Finally, we migrate the modified parameters into the generated tests, resulting in a 63.3\% improvement in performance.

We use static analysis tools (e.g., javacg-static) to locate the entry function and vulnerable function, which guides generating a high test coverage. If the test case successfully reproduces the vulnerability after incorporating the parameters, the project is deemed to have an exploitable library vulnerability, and the test case serves as the exploit in the project. Additionally, though projects' existing tests are not effective in detecting potential library vulnerabilities ~\cite{Iannone1}, they provide extensive information related to the function call in the project. If existing tests cover the vulnerable function, we can migrate parameters into these tests instead of generating tests.

%讲述实验结果
We evaluate \appname and our baseline, \baselinename, by using a dataset consisting of 30 vulnerabilities and 60 projects affected by library vulnerabilities sourced from GitHub. \appname successfully generates 43 exploits for 26 vulnerabilities, outperforming \baselinename, which produces 11 exploits. On average, it takes 22.75 seconds to generate an exploit when no existing test case is available. Additionally, we test our method in eight projects with tests covering vulnerable functions. By migrating parameter values into existing tests, all of the projects' library vulnerabilities are exploited.

%讲述贡献
In summary, we make the following contributions: 
\begin{itemize}[leftmargin=*]
\item {We propose a migration-based test generation method to provide domain knowledge for triggering the vulnerability. Our tool is available on our website\footnote{https://github.com/chen-zirui/TestMigration}.}

\item {We implement an exploit generator that exploits library vulnerabilities by utilizing the vulnerable function's position within the project and domain knowledge provided by library exploit code.}

\item {We evaluate our method on 30 known vulnerabilities across 60 Java vulnerability-project pairs, resulting in the successful generation of 32 additional exploits compared to baseline \baselinename.}
\end{itemize}

%文章展开
The remainder of the paper is organized as follows. In Section 2, we present a motivating example. In Section 3, we describe the implementation details of our method. In Section 4, we demonstrate the effectiveness of our approach through empirical evaluation. In Section 5, we discuss the reliability of our method. In Section 6, we provide an overview of the related work in this paper. Finally, in Section 7, we summarize our method and mention future works.
\section{Motivation Example}

\begin{figure}[htbp]
    \centerline{\includegraphics[width=8cm]{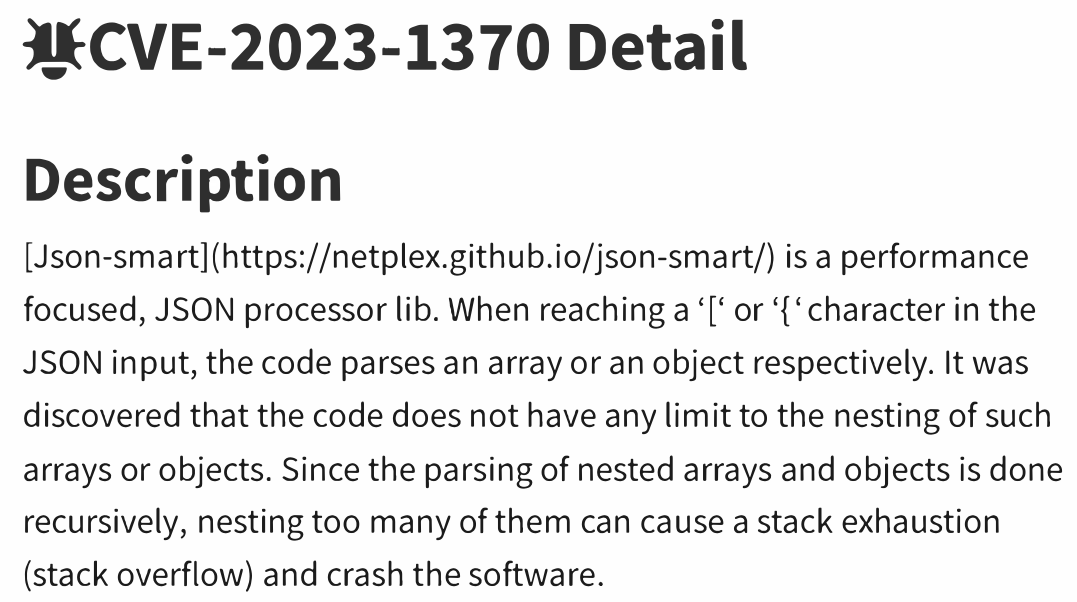}} %添加图片路径
    \caption{Description of CVE-2023-1370.} %图片描述
    \label{fig:1370} %图片索引
\end{figure}

%motivation example
%漏洞通报包含的信息与漏洞通报无法包含的东西
Figure \ref{fig:1370} presents the description of CVE-2023-1370, which provides information about the vulnerability, e.g. the root cause and the reproduction steps of the vulnerability. However, the description only mentions that a processed input stream will result in stack overflow without providing specific information about the details of the input. The lack of details makes it difficult for developers to determine whether the project has been affected as developers don't know which input will cause the problem.

%漏洞利用包含的信息, 以及开发者应该如何使用
As shown in Figure \ref{fig:poc-1370}, in some security research websites \footnote{https://research.jfrog.com/}, we can find the vulnerability exploit (POC) for CVE-2023-1370. The exploit initially constructs a string with specific characteristics to trigger the vulnerability, which uses two \texttt{for} loops to make a nested object. Subsequently, this value is passed to the vulnerable function. Executing this exploit code on a vulnerable version of the library will result in a denial of service issue. This exploit reveals that if the vulnerable function receives a value similar to $s$, the project will encounter a stack overflow problem. To identify the vulnerability in the project, developers must search for all potential entries of the vulnerable function and verify if users can create an input resembling $s$.

\begin{figure}[ht]
    \centering
    \begin{lstlisting}[language=Java,xleftmargin=1em]
StringBuilder s = new StringBuilder();
for (int i = 0; i < 10000 ; i++) {
  s.append("{\"a\":");
}
s.append("1");
for (int i = 0; i < 10000 ; i++) {
  s.append("}");
}
JSONParser p = null;
p = new JSONParser(JSONParser.MODE_JSON_SIMPLE);
p.parse(s.toString());
    \end{lstlisting}
  \caption{Vulnerability exploit for CVE-2023-1370 in jfrog.}

\label{fig:poc-1370}
\end{figure}

%通过方法生成的测试与帮助开发者
Our objective is to identify vulnerabilities associated with CVE-2023-1370 in an open-source project named \textit{microservice-with-jwt-and-microprofile}. Within the project, there is a code snippet that invokes \texttt{parse} as follows: \texttt{parser.parse(content)}. Given the project with the exploit of CVE-2023-1370, \appname generates some test cases to trigger the vulnerability as clients of the project. Figure \ref{fig:test} presents the test case generated for this vulnerability, with line 4 denoting the invocation of the entry function. Replacing the parameter with a specific value such as the aforementioned $s$ will result in the execution of this test case throwing an uncaught exception due to a buffer overflow problem. For projects with existing tests, \appname will try to generate an exploit based on the graph from the test to the vulnerable function by replacing a triggerable value into the test directly.

%给一个基于现有测试的例
With the assistance of \appname, developers can execute the generated test case to ascertain if users can construct an input that triggers a buffer overflow in the project. For instance, executing \texttt{test05} in Figure \ref{fig:test} will cause result in a denial-of-service in the project, which means developers need to update the dependency version. Our method prevents false alarms by using vulnerability triggers as reliable evidence. Furthermore, we offer developers trigger test cases to assist in vulnerability identification in their projects and gain insight into how attackers may exploit the vulnerability.

\begin{figure}[ht]
    \centering
    \begin{lstlisting}[language=Java,xleftmargin=1em]
@Test(timeout = 4000)
public void test05()  throws Throwable  {
  try { 
    TokenUtil.of(POCValue);
    fail("Expecting exception: ClassCastException");
  } catch(ClassCastException e) {
     verifyException("TokenUtil",e);
  }
}
    \end{lstlisting}
  \caption{The generated test case for exploiting CVE-2023-1370 in the project named Microservice.}

\label{fig:test}
\end{figure}

\begin{figure*}[htbp] 
  \centering	
  \includegraphics[width=14.5cm,scale=0.90]{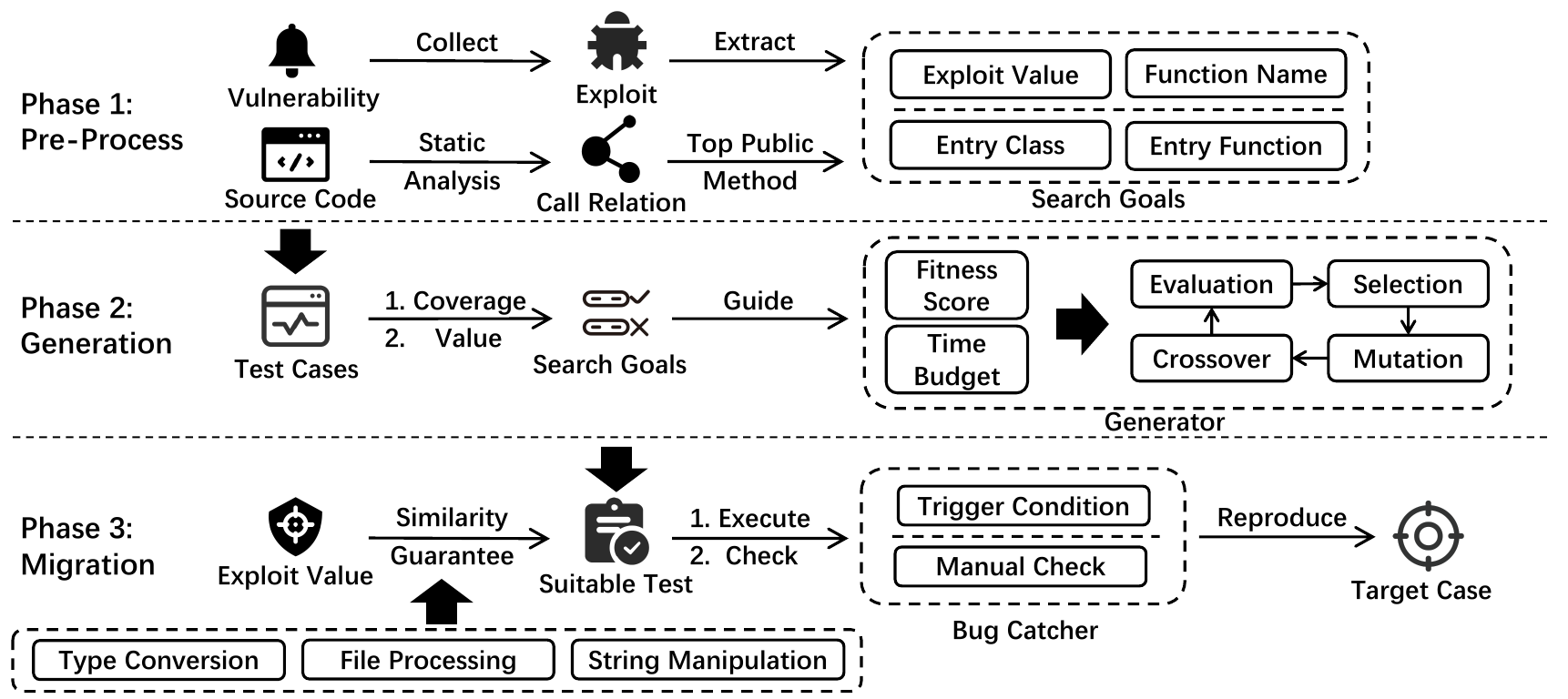}
  \caption{Overview of VESTA. Given the vulnerability exploits code, VESTA produces an exploit (Test Case) for the potential library vulnerability in the project.}
  \label{fig:FigureOne}
\end{figure*}

\section{Proposed Approach}
To generate test cases for triggering library vulnerabilities in projects, we implement a method called \appname, as shown in Figure \ref{fig:FigureOne}. Giving the binary file of the project and the exploit code for the library vulnerability, \appname initially analyzes line coverage goals associated with the vulnerable function and assesses if the project's existing tests contain a call graph to the target function. Subsequently, \appname generates test cases using a genetic algorithm to achieve the highest possibility to call the vulnerable function. \appname generates tests that differ from those generated by EvoSuite in that they include call graphs passing parameters to the vulnerable function. These test cases are executed with an instrumentor and call graphs containing the vulnerable function are collected. Lastly, \appname integrates the exploit code into the call graphs and assesses if the vulnerability is reproduced when executing the modified test cases.

\subsection{Pre-Processing}
\label{sec:pre-process}
\appname takes the target project, along with its existing tests, as input. The pre-processing step has two major objectives: 1) To collect search goals for the generation step, including identifying the class name where the entry function is located and the position of the vulnerable function in the project. 2) To generate dynamic call graphs of the project's tests. If the analysis is based on the existing tests of the project, we compile and execute the current tests to obtain the dynamic invocation chain. If the analysis is based on the generated tests, we perform instrumentation analysis using the runtime package of EvoSuite.

To identify the search goals, we utilize a static analysis tool called \textit{javacg-static} from Github\footnote{https://github.com/gousiosg/java-callgraph}. The entry function is determined as the public method at the top of the static call graph containing the vulnerable function. The call graphs are obtained using a depth-first algorithm, which analyzes the function call relations collected by static analysis.

\begin{algorithm}[htbp]
    \caption{Test Execution and Instrument}
    \label{alg:algorithm1}
  
    \SetKwInOut{Input}{Input}
    \SetKwInOut{Output}{Output}
    \Input{Test case files $test\_files$, target function $target\_function$}
    \Output{Dynamic call graph $call\_graph$}
    \For{$test$ \textbf{in} $test\_files$}{
        \If{$vulnerable\_function$ \textbf{in} $test$}{
            execute $test$ \\
            \For{$method$ \textbf{in} $test$}{
                $insertPushStatement(method)$ \\
                \If{$method$ \textbf{equals} $target\_function$}{
                    $call\_graph \gets$ $getCallGraph()$
                }
                $execute(method)$ \\
                $insertPopStatement$
            }
        }
    }
 
\end{algorithm}

As shown in Algorithm \ref{alg:algorithm1}, We design an instrumentor to collect the dynamic call graph during the test execution. The instrumentor inserts a push operator (line 5) before each function call and a pop operator (line 9) after each function call when executing tests. When the target function is called, the dynamic call graph containing the target function is stored on the stack (line 7).

\subsection{Test Generation}
After pre-processing, if no existing test is suitable for exploit migration, \appname will generate test cases for the project, which contains a call graph from the entry to the vulnerable function. Due to the maturity of EvoSuite ~\cite{Almasi1, Devroey1}, \appname reuses EvoSuite's genetic algorithm implementation.

\textbf{Search Goals.} In the migration step ~\ref{sec:migration}, \appname selects a test case covering the target function and modifies the test to create an input that triggers the vulnerability in the vulnerable function. To achieve this objective, our search goals consist of two parts: 1) Ensuring that the generated test covers the vulnerable function. 2) Allowing the input value passed to the vulnerable function to be modified by adjusting the value provided in the generated test.

Generating a test case that invokes the vulnerable function can be challenging. To generate high coverage test cases for vulnerable functions in the project, \appname collects three pieces of information as search goals: 1) \textit{Entry Class Name} guides EvoSuite in generating tests for the function in the entry class, which represents how the client will call the function; 2) \textit{Entry Function Name} is used to guide EvoSuite generating tests that contain the user entry function in the project, avoiding directly call the vulnerable function; 3) \textit{Vulnerable Function Name} is used to verify whether the generated test can reach the vulnerable function. \textit{Entry Class Name} and \textit{Entry Function Name} are collected by our pre-processing step while \textit{Vulnerable Function Name} is manually collected from the publicly disclosed vulnerabilities databases, like CVE ~\cite{CVE1}.

To ensure that the generated test can pass values to the target function, we establish a search objective linked to the parameter value transmitted to the vulnerable function. We compare this value with the value in the exploit code, and if it changes during the test generation step, we can confirm the existence of a call graph that can provide crafted input to the vulnerable function. In some instances, this search goal leads to generating test cases that trigger the vulnerability directly.

\textbf{Fitness Function.} 
To generate tests that meet our search goals, we adjusted the fitness function in EvoSuite. As demonstrated in Algorithm \ref{alg:algorithm2}, \appname's fitness function assesses the proximity of the generated test to reach the library vulnerability function. The line coverage score measures the proximity of the test to accessing the vulnerable function, while the parameter similarity score evaluates the feasibility of passing the parameter from the test and triggering the vulnerability.

\begin{algorithm}[htbp]
  \caption{Fitness Function for each Generation}
  \label{alg:algorithm2}
  
    \SetKwInOut{Input}{Input}
    \SetKwInOut{Output}{Output}
    
    \Input{The coverage goal $goal$, exploit value $poc$, generation $gen$}
    \Output{Individual fitness score $score$}
    \For{$ind$ \textbf{in} $gen$}{
        $(entry, function, value) \gets execute(ind)$  \\
        $score \gets 0$ \\
        \If{$entry$ \textbf{in} $goal$}{
            $score \gets score + 1$
        }
        \If{$function$ \textbf{in} $goal$}{
            $score \gets score + 1$ 
        }
        \Switch{$value$}{
            \Case{string}{
                $score \gets score + distanceScore(value, poc)$ 
            }
            \Case{number}{
                $score \gets score + isSameValue(value, poc)$ 
            }
            \Case{object}{
                $score \gets score + inspectorScore(value, poc)$ 
            }
            \Case{file}{
                $obj \gets convertToFileObject(value)$ \\
                $score \gets score + inspectorScore(obj, poc)$
            }
        }
    \Return{$score$}
    }
\end{algorithm}

In \appname, line coverage employs branch distance as a guiding metric for test generation. The branch distance helps generate tests that trigger vulnerable functions by measuring the proximity between the vulnerable function's branch and the actual branch taken by the generated test case.

To account for parameter similarity in the fitness score, we compute the similarity between the actual parameter value that the vulnerable function gets when executing the test and the value in the exploit based on the parameter types. Our work includes four types: number, string, file, and object.

{\bf Number. } In Java, the number type includes Int, Long, Short, Double, Float, char, and Boolean. Number type parameters are compared directly with the value in the exploit code. If the value in the generated test matches the value in the exploit, the similarity is 1, otherwise, the similarity is 0.

{\bf String. } We use edit distance ~\cite{Parsing1} to evaluate the similarity between the actual and exploit parameters, which represents the number of steps required to change one string to the target string. To calculate similarity, \appname divides the edit distance by the length of the longer string and subtracts the result from 1. If the two strings are identical, the similarity is 1.

{\bf Object. } \appname calculates the similarity of two objects by the average similarity between inspectors ~\cite{Hong1, Dallmeier1}. The corresponding similarity calculating methods are selected by the inspectors' type.

{\bf File. } Files are treated as objects in \appname due to their representation as File objects in Java. The exploit parameter is stored as a position on the disk. During the comparison, \appname retrieves the exploit file from the disk and compares it with the actual value passed to the vulnerable function. 

\textbf{Implementation.} 
We use javacg-static to get the location of the vulnerable function within the project's binary file. Our approach utilizes the existing infrastructure of EvoSuite 1.0.4 and involves modifying the fitness function within EvoSuite. The coverage of the vulnerable function is measured using branch distance. The similarity is assessed based on the parameter types of the vulnerable function, ensuring the presence of the call graph from the test to the vulnerable function. We use Javassist\footnote{https://www.javassist.org} to design our instrumentor, which collects dynamic call graphs during test execution and replaces entry parameter values while migrating.

\subsection{Exploit Migration}
\label{sec:migration}
\appname includes a migration step in the generated test to trigger the vulnerability and significantly improves effectiveness by utilizing domain knowledge extracted from vulnerability exploits. Instead of using the triggering condition, we use the parameter value passed to the vulnerable function in the exploit to guide test generation. Additionally, rules are collected from the situations in which the parameters are modified in the process of passing value from the entry function to the vulnerable function.

\textbf{Exploit Extraction.} Triggering library vulnerabilities in a project is typically manifested as calls to vulnerable APIs. For example, Figure \ref{fig:poc} illustrates the exploit for CVE-2021-44248, which invokes the \texttt{logger.error} function with a specific value. Executing the exploit triggers remote code execution caused by the Apache-log4j2\footnote{https://logging.apache.org/log4j/2.x/} library. The vulnerability exploit contains information about the value passed to the vulnerable function, which is treated as domain knowledge for reproducing the library vulnerability. In \appname, we collect the parameter value passed to the vulnerable function in the exploit by executing the exploit code snippets to ensure generate a test case that passes a similar value as the value in the exploit code.

\begin{figure}[ht]
    \centering
    \begin{lstlisting}[language=Java,xleftmargin=1em]
private static final Logger logger = LogManager.getLogger(log4j.class);
public static void main(String[] args) {
    logger.error("${jndi:ldap://localhost:8080/exploit}");
}
    \end{lstlisting}
  \caption{The exploit code of CVE-2021-44248, will cause the exploit.class to run on the server (in the exploit, it is hosted on localhost).}

\label{fig:poc}
\end{figure}

\textbf{Parameter Migration.} \appname utilizes the parameter value extracted from the exploit code to trigger the library vulnerability during the processing of the generated test cases. The fitness function of the genetic algorithm incorporates a calculation to assess the similarity between the test's passing value and the parameter value of the exploit. A test attains a high fitness score when a call graph that enables the parameter value to be passed from the test to the vulnerable function exists. Given the presence of this call graph, modifying the parameter at the entry of the call graph (Test) could alter the value received by the vulnerable function at the end of the call graph.

Instrumentor is a crucial component in \appname which enables dynamic analysis and monitoring of program execution. It collects runtime information, such as dynamic call graphs, and allows for the modification of parameter values during execution.

For a specific vulnerability, \appname retrieves triggering parameter values from the exploit. By executing the generated test cases, we can obtain the entry function during testing. \appname then re-executes the test with the instrumentor, which replaces the entry function's parameter value and records the target function's received value. When the executing entry function, its parameter will be replaced with the retrieved value and the parameter passed to the vulnerable function is recorded. If the entry function has more than one parameter, \appname will traverse all positions and obtain the correct position to pass the value to the target function.

\textbf{Trigger Capture.} 
For each vulnerability type, \appname defines the trigger condition of the vulnerability. Our experiment part includes these types: Denial of Service (DOS), Remote Code Execution (RCE), Function Wrong Behavior, SQL Injection, and XML External Entity Injection (XXE). Table \ref{tab:conditions} shows the definition of the conditions in \appname. During the test execution process, \appname collects the test execution conditions and checks if they meet our defined criteria. A test that matches the defined criteria is considered an exploit test for the library vulnerability in the project.

However, manual confirmation is still necessary to verify some vulnerabilities for the following reasons: (1) After fixing the vulnerability, the only observable change may be in the API return value without any other behaviors such as exceptions; (2) EvoSuite's sandbox implementation limits file access, so developers should log the return value of the vulnerable function or manually execute the generated test to check for trigger conditions.

\textbf{Similarity Guarantee.} 
Under certain conditions, solely replacing the value of the entry function may not trigger the vulnerability due to modifications within the call graph parameters. To overcome this issue, it is necessary to assess the potential for reproducing the vulnerability and modify our primitive value accordingly. By executing the migrated test and monitoring code conditions, \appname assesses the possibility of the test triggering the vulnerability. If the test fails to reproduce the library vulnerability, \appname utilizes collected rules to process the migrating parameter to ensure the similarity between generated test and the exploit.

As shown in Algorithm \ref{alg:algorithm3}, during the execution of the test process, if the test fails to directly trigger the vulnerability, \appname attempts to utilize rules to manipulate the parameter values within the test's entry function and re-execute the test, aiming to trigger the vulnerability. Typical rules encompass parameter type conversion, string manipulation, and file processing.

\begin{table}[t]
    \centering
    \caption{Trigger conditions of vulnerabilities in VESTA.}
    \label{tab:conditions}
    \begin{tabular}{lcccrrr}
    \toprule
        \textbf{Type} & \textbf{Show}  \\ \midrule
         DOS & Uncatch exception / Infinite loop \\
         RCE & Target server receive request     \\
         Wrong Behavior & No exception throwed   \\
         SQL Injection  & Database unexpected logs  \\
         XXE & Target server receive request     \\
    \bottomrule
    \end{tabular}
\end{table}

\textit{Parameter Type Conversion.} The parameter type is usually converted from the input to the vulnerable function during the parameter transfer process. For example, the target function expects a parameter of type \texttt{ByteArray}, whereas the entry function requires a parameter of type \texttt{String}. During the migration process, \appname obtains the parameter type from the exploit (Usually \texttt{String}), as well as the expected type of the entry function from the static analysis results. It then modifies the input type accordingly. 

\textit{String Manipulation.} String manipulation refers to the process of altering a String-like exploit value by appending a substring and incorporating it into an object or another format. Adding a substring involves inserting a string into specific locations within the value or incorporating the value into a predefined template string (e.g., inserting the server IP into an exploit string in RCE vulnerabilities). In some cases, the input value may represent only a portion of a specific format, such as a value within an object type. In such situations, the exploit string is inserted into an incomplete object or JSON template.

\textit{File Processing.} Certain vulnerability exploits involve specific files that trigger vulnerabilities, and these files are stored at specific disk locations. In genetic algorithms, generating files with such specific characteristics can be challenging, even with knowledge of the file processing rules. To reduce the generation number and improve the correct rate, rather than collecting rules for file processing, we streamline the migration process by directly converting file positions into corresponding file objects in the executing process. \appname accesses and converts these files into file objects during migration to facilitate subsequent processing.

\begin{algorithm}[htbp]
    \caption{Modifying Parameter Value}
    \label{alg:algorithm3}
    
    \SetKwInOut{Input}{Input}
    \SetKwInOut{Output}{Output}
    \Input{Generated test $generated\_test$}
    \Output{Modified test $modified\_test$}
    $modified\_test \gets generated\_test$; \\
    \If{$generated\_test$ not \textbf{in} $trigger\_vulnerability$}{
        \If{$parameter\_type$ not \textbf{equals} $trigger\_type$}{
            $generated\_test \gets typeConvertion(generated\_test)$;
        }
        $generated\_test \gets stringManipulation$; \\
        $reRexecution(generated\_test)$;
    }
    \Return $generated\_test$;

\end{algorithm}

\textbf{Exploit Generation.} Following parameter migration, \appname re-executes the migrated test. If the vulnerability trigger condition is satisfied, the test is reported as an exploit for the library vulnerability in the project. To better understand the vulnerability, the call graph from the entry function to the vulnerable function is provided. For projects without library vulnerability, \appname generates tests equally but the generated tests are unable to reproduce the vulnerability.  

Additionally, during the static analysis process (pre-process in Section \ref{sec:pre-process}), if the project's existing tests contain a call graph to take into the vulnerable function, \appname will skip the test generation step and solely migrate the exploit parameter to the test. If the existing test fails to identify a library vulnerability, we will generate tests for the project as projects without satisfying tests. 
\section{Experiments}

% Table generated by Excel2LaTeX from sheet 'pocdata'
\begin{table*}[htbp]
  \centering
  \caption{Vulnerabilities in our experiment, including vulnerability numbers, library names, vulnerability trigger conditions. Each Vulnerability has two projects, we indicate \ding{51}  if the method is a success in the project, \ding{55} if the method failed in the project, \textbf{——} means no vulnerability witness test is added after fixed or the fixing test is not merged.}.
  \label{tab:result}
    \begin{tabular}{lllllcc}
    \toprule
    \multicolumn{1}{l}{\textbf{Type}} & \textbf{Library} & \textbf{Number} & \textbf{Function} & \textbf{Trigger Condition} & \textbf{VESTA} & \multicolumn{1}{l}{\textbf{TRANSFER}} \\
    \midrule
    Xml   & XStream  & CVE-2017-7957 & fromXML & Uncatch Exception & \ding{51} \ding{51}   &  \ding{51} \ding{51}\\
          &       & CVE-2021-39144 & fromXML & Remote Code Execution & \ding{51} \ding{51}    &   \textbf{——}  \\
          &       & CVE-2021-21341 & fromXML & Infinted Loop & \ding{51} \ding{51}    &  \textbf{——}  \\
          &       & CVE-2022-41966 & fromXML & Stack Overflow & \ding{51} \ding{51}   &  \ding{51} \ding{51}\\
          &       & CVE-2020-26217 & fromXML & Remote Code Execution & \ding{51} \ding{51}   & \ding{55} \ding{55}  \\
    \midrule
    Base64/32 & Apache Codec & CODEC-263 & decodeBase64 & Wrong Behavior & \ding{51} \ding{55}    &            \textbf{——}  \\
          &       & CODEC-270 & decodeBase64 & Wrong Behavior & \ding{51} \ding{51}    &  \ding{51} \ding{51}  \\
    \midrule
    String & Apache Text & TEXT-215 & translate & Wrong Behavior & \ding{51} \ding{51}    & \textbf{——}  \\
          &       & CVE-2022-42889 & replace & Remote Code Execution & \ding{51} \ding{55}    & \ding{55} \ding{55}  \\
    \midrule
    Number & Apache Lang & LANG-1484 & isParsable & Wrong Behavior & \ding{51} \ding{55}    &  \textbf{——}  \\
          &       & LANG-1645 & createNumber & Wrong Behavior & \ding{51} \ding{51}    & \ding{55} \ding{55}  \\
          &       & LANG-1385 & createNumber & Wrong Behavior & \ding{51} \ding{51}    & 
          \ding{55} \ding{55} \\
    \midrule
    Json & Json-smart & CVE-2023-1370 & parse & Stack Overflow & \ding{51} \ding{55}    &   \ding{55} \ding{55} \\
          &       & CVE-2021-27568 & getAsNumber & Uncatch Exception &  \ding{51} \ding{55}   &   \ding{51} \ding{55}  \\
          & JSON  & CVE-2022-45688 & parse & Uncatch Exception & \ding{51} \ding{51}   &  \ding{51} \ding{55}  \\
          & Jackson-databind  & CVE-2019-14540 & readValue & Remote Code Execution & \ding{55} \ding{55} &  \ding{55} \ding{55} \\
    \midrule
    File & Apache.poi & CVE-2019-12415 & XSSFExportToXml & XXE Injection &                \ding{51} \ding{51}    & \textbf{——}  \\
          & Zip4j & CVE-2022-24615 & ZipInputStream & Uncatch Exception & \ding{51} \ding{55}    &  \textbf{——}\\
          &       & Zip-263 & ZipFile & Wrong Behavior & \ding{51} \ding{51}    & \ding{51} \ding{51}  \\
          & Apache IO & IO-611 & normalize & Path Traversal & \ding{51} \ding{51}    & \ding{55} \ding{55} \\
          &       & CVE-2021-29425 & normalize & Path Traversal & \ding{51} \ding{51}    & \ding{55} \ding{55} \\
          & Apache PDFBox & CVE-2021-31812 & load & Stack Overflow  & \ding{51} \ding{51}    &  \textbf{——} \\
    \midrule
    Compress & Apache Compress & CVE-2021-35516 & SevenZFile & Out of Memory         &  \ding{51} \ding{55}    &   \ding{51} \ding{55}\\
            &       & CVE-2018-1324 & ZipFile   & Wrong behavior   &  \ding{51} \ding{51}  & \textbf{——}\\
    \midrule
    Test  &  Junit   &   CVE-2020-15250 &  TemporaryFolder  & Improper File Permission  &                \ding{55} \ding{55}  &  \ding{55} \ding{55}\\
    \midrule
    Framework  &   Spring-beans   & CVE-2022-22965 &  ClassLoader  & Remote Code        Execution &  \ding{55} \ding{55}  & \textbf{——} \\
    \midrule
    Net  &  Httpclient   &   CVE-2020-13956 &  HttpGet  & Cross-site Scripting  &                \ding{51} \ding{51}  & \ding{55} \ding{55}\\
    \midrule
    HTML  &   Jsoup  &   CVE-2021-37714  &   parse   &  Infinted Loop  &         \ding{51} \ding{55}     &   \ding{55} \ding{55}  \\ 
    \midrule
    Log  &  Log4j2   &   CVE-2021-44228 &  error/info  & Remote Code Execution &           \ding{51} \ding{55}      &  \ding{55} \ding{55} \\
    \midrule
    Database  &  Hibernate   &   CVE-2019-14900 &  getResultList  & SQL injection &             \ding{55} \ding{55}  &  \ding{55} \ding{55}  \\
    \midrule
              &  18 Libraries   &   30 Vulnerabilities &    &  &   43/60   & 11/60\\
        
    \bottomrule
    \end{tabular}%
\end{table*}%

We conduct an empirical evaluation to assess the effectiveness of our method in detecting library vulnerabilities. The evaluation aims to answer the following research questions:

\begin{itemize}[leftmargin=*]
\item \textbf{RQ1. Is VESTA effective in generating exploits for library vulnerabilities? }

This question focuses on assessing the effectiveness of \appname in vulnerability discovery. We evaluate the accuracy of \appname in a manually selected vulnerability dataset. \baselinename ~\cite{Hong1} is used as the baseline for comparison to determine if both approaches can identify exploitable vulnerabilities and assess the time cost for their discovery.
\item \textbf{RQ2. Does exploit migration enhance the effectiveness of VESTA?}

The experimental section will discuss the performance of vulnerability detection based on migration compared to direct test generation, aiming to validate the advantages of exploit generation with our migration step.
\end{itemize}

\subsection{Experimental Setup} 
Here, we discuss the experimental subjects used in our study, including the collected dataset and our baseline.

\textbf{Context Selection.}The experimental section analyzes 30 reported vulnerabilities from the past 5 years. Two projects associated with the corresponding library and affected by the vulnerabilities are selected for experimentation for each vulnerability. We exclude toy projects from our dataset and only select those with 1000 or more code lines. We have a total of 41 projects, and we provide the complete list of projects on our website. To ensure the generalizability of the method, the dataset includes various types of vulnerability types, such as Denial of Service (Infinite Loops, Uncaught Exceptions), Wrong Function Results, Remote Code Execution, XML Data Injection, and SQL Injection. Additionally, experiments are conducted on diverse types of libraries with different functionalities, such as JSON processing, Java testing frameworks, Base64 conversion, and HTTP frameworks. Under this criterion, the vulnerabilities involved in the experiments are not limited to a single domain or a specific type. As Table \ref{tab:type}, our experiment includes four common vulnerable parameter types, which manifests that our method is effective for different API-level vulnerabilities. We categorize the various conditions of vulnerability triggers into five types, which aim at avoiding manual checking of triggers. These types cover 8 of the 10 most common CWE in 2022  ~\cite{Mend1}, with CSRF and XSS being the only exceptions.

\begin{table}[t]
    \centering
    \caption{Vulnerable function parameter types in experiment.}
    \label{tab:type}
    \begin{tabular}{cc}
    \toprule
        \textbf{Parameter Type} & \textbf{Vulnerability} \\ \midrule
         String & 20 \\
         File &  3 \\
         Object & 5  \\
         Number & 2 \\
    \bottomrule
    \end{tabular}
\end{table}

For each vulnerability, the dataset includes two projects that depended on libraries affected by that vulnerability and one project did not affected by the vulnerability although exists the vulnerable function call. In order to compile these projects successfully and obtain more experimental data, we make certain modifications, such as changing dependency versions and removing uncompilable files, while avoiding any modifications to functions present in the vulnerable function call graphs to prevent false positives. Exploit code is selected from the report of each vulnerability, and the corresponding exploit parameters are extracted. Projects involved in experiments are Java projects managed by Maven.

To verify the feasibility of vulnerability exploit migration based on existing tests, we select four vulnerabilities and each two corresponding projects with comprehensive tests during project collection. We conduct experiments on the aforementioned vulnerabilities to validate the rationale of utilizing existing tests for discovering exploitable vulnerabilities. Moreover, the experimental section also designs a comparison between vulnerability discovery based on test generation and based on existing tests on the same projects to discuss the similarity between generated tests and the manually designed test. The vulnerabilities involved in the experiments are presented in Table \ref{tab:result}.

During the experiment setup, vulnerability information is obtained from CVE or the libraries' vulnerability reports. The exploit code used in the experiment is sourced from various open-source forums (e.g., snyk ~\cite{Snyk1}), and the vulnerable projects are sourced from GitHub, ensuring the authenticity of the experiment.

\textbf{Baseline.} We compare our method with \baselinename~\cite{Hong1}, which generates tests for exploiting library vulnerabilities guided by the code behavior captured while executing vulnerability-witnessing tests. To run \baselinename in our experiment, we collect positions of vulnerable functions within the projects and vulnerability-witness tests from vulnerable libraries. We evaluate \baselinename in our experiments based on the example provided in the code package of \baselinename. 

\textbf{Setup.} 
To determine if a test triggers the vulnerability, \appname captures the execution result of the generated tests and checks if a triggering behavior, corresponding to our defined trigger condition, has occurred. A test that exhibits the triggering behavior during execution is considered an exploit. 

A test case is classified as an exploit if a defined trigger condition occurs and is detected by \appname during test execution. Both methods will be executed 10 times for each project, and the method is considered to be effective if generates the exploit 5 times or more. However, if the method falsely determines a project without exploitable vulnerabilities as positive, it will be marked as a false positive.

We conduct our experiments on a 3.5GHz M2 device with 16GB RAM. Following Kang et al.'s experimental result ~\cite{Hong1}, we set the test generation time budget in both approaches for 60 seconds.

\subsection{Results}
Here, we answer our two research questions by analyzing the experiment results.

\subsubsection{RQ1: Effectiveness of \appname} 

Table \ref{tab:result} presents the results of \appname's detection of vulnerabilities. In projects containing exploitable vulnerabilities, \appname identified 71.7\% (43/60) of exploitable vulnerabilities while on the same dataset, \baselinename only confirmed 18.3\% (11/60) of exploitable vulnerabilities. \appname outperformed \baselinename  on 20 vulnerabilities. This result demonstrates the effectiveness of \appname in identifying exploitable library vulnerabilities and generating corresponding exploit code.

Our experiment evaluates the average time taken by \appname for generating exploitable vulnerabilities. Based on existing tests, \appname only requires preprocessing to obtain potential vulnerability call graphs and performs the migration step to generate an exploit. For projects without existing tests, the time search budget for the generated work during test generation is uniformly set to 10 seconds. In the experiments, all successful cases completed the test generation step within 10 seconds, demonstrating its superior performance in practice. On projects with tests containing exploitable call graphs, \appname takes an average of 3.09 seconds to discover exploitable vulnerabilities. On projects without complete test coverage, the average time for discovering exploitable vulnerabilities is 22.75 seconds.

We run \appname on projects with exploitable vulnerability call graphs that are unable to trigger vulnerabilities, causing false alarms in traditional vulnerability discovery methods. These projects are often mistakenly identified as having exploitable vulnerabilities in traditional vulnerability discovery methods like dependency analysis, leading to false alarms. In all 30 projects, \appname consistently determines the absence of exploitable vulnerabilities, which aligns with the expected outcome.

\textbf{Answer to RQ1.} \appname can generate exploit code for 43 projects associated with 26 third-party vulnerabilities, whereas the baseline \baselinename can only generate exploit code for 11 projects. Furthermore, \appname does not produce any false alarms, thereby demonstrating its reliability in the task of project vulnerability discovery.

\subsubsection{RQ2: Ablation Study} 

We conduct an ablation study on \appname by removing components from it. The results of this experiment are shown in Table \ref{tab:ablation}. Compared to EvoSuite, \appname incorporates vulnerability-driven migration tasks. Without migration tasks, \appname can only generate 5 exploit tests related to 3 vulnerabilities (CODEC-263,  CODEC-270, Zip-263), which aligns its performance with EvoSuite configured with line coverage and branch coverage. After incorporating the migration step, \appname can generate exploits for an additional 38 projects, demonstrating the effectiveness of \appname's vulnerability discovery based on migration.

Table \ref{tab:complete} presents the ablation study conducted on projects with complete tests. Experiments are conducted on these projects using the following methods: 1. Migration using existing tests, 2. Migration using generated tests, and 3. Experimentation using only generated tests. This experiment demonstrates that migration based on existing tests and migration based on generated tests both exhibit good performance, while direct generation performs poorly.

\begin{table}[t]
    \centering
    \caption{Results of Ablation Study on VESTA.}
    \label{tab:ablation}
    \begin{tabular}{ccc}
    \toprule
        \textbf{Method} & \textbf{Exploit}   & \textbf{Effectiveness}\\ \midrule
         Migration with Rules      & 11 &   18.3\% \\
         Directly Migration        & 27 &   45.0\%   \\
         Directly Generated Test   & 5 &    8.3\% \\
         Total  & 43 & 71.6\% \\
    \bottomrule
    \end{tabular}
\end{table}

\begin{table}[t]
    \centering
    \caption{VESTA's performance on projects with complete tests in three scenarios: direct migration, migration after generation, and test generation only.}
    \label{tab:complete}
    \begin{tabular}{cccc}
    \toprule
        \textbf{Method} & \textbf{Found} & \textbf{Omitted} & \textbf{accuracy}\\ \midrule
         Migration on Generated Test & 8 & 0  & 100.0\%   \\
         Migration on Existing Test   & 8 & 0    & 100.0\%   \\
         Directly Generated Test   & 4 & 4 & 50.0\% \\
    \bottomrule
    \end{tabular}
\end{table}

Table \ref{tab:time} compares the time required for generating exploit code on the vulnerabilities in which both \appname and \baselinename have successful cases. Since \baselinename relies on manual confirmation for locating the vulnerable function positions in the projects, the experiment omits the time spent on this part and focuses only on comparing the generation time in \baselinename with the generation and migration time in \appname. This experiment revealed that when complete tests already exist for the projects, \appname can rapidly generate exploit code based on the existing project tests. Furthermore, the time performance for migration based on generated tests is also better than the direct generation. 

\begin{table}[t]
    \centering
    \caption{Exploit Generation Time. In VESTA, we evaluate time both using existing tests and generating tests.}
    \label{tab:time}
    \begin{tabular}{lcc}
    \toprule
        \textbf{Vulnerability} & \textbf{VESTA} & \textbf{TRANSFER} \\ \midrule
         CVE-2017-7957  & 2.20s/18.71s &  31s   \\
         CVE-2022-41966 & 2.51s/18.41s &  32s   \\
         CODEC-270      & 3.51s/16.33s &  12s   \\
    \bottomrule
    \end{tabular}
\end{table}

\textbf{Answer to RQ2.} \appname's migration part achieves a success rate of 63.3\% and showed an average improvement of 7.81 seconds in time compared to \baselinename. Therefore, the approach of migration based on test generation exhibits efficiency and reliability in discovering third-party exploitable vulnerabilities in projects.

\section{Discussion}
In this part, we discuss the reliability and threats to validity of our study.

\subsection{Qualitative Analysis}
The fitness function of \baselinename in test generation aims to ensure the generated tests exhibit similar code behavior as the tests added to the vulnerable repository after fixed. However, this approach presents several issues. Firstly, some vulnerabilities, such as CVE-2022-24615, do not have additional tests added after the vulnerability is repaired (11 vulnerabilities). Meanwhile, a no runnable methods exception will occur if the libraries' test framework is JUnit5 instead of JUnit4 (8 vulnerabilities). Another problem arises when the vulnerability repository tests invoke methods specific to the test classes, which cannot be called in the target project, as exemplified by CVE-2020-26217. Lastly, tests generated based on code behavior still exhibit differences compared to actual tests capable of triggering vulnerabilities, particularly when file operations are involved, as demonstrated by CVE-2019-12415.

In contrast, \appname exploits vulnerability by generating tests and obtaining exploitable call graphs, ensuring that all methods called in the tests are existing methods within the project. In the experiment involving CVE-2020-26217, \baselinename utilized a code remote execution detection method from the XStream library test. However, since developers require to execute the test within their projects, the relevant files for this method are not included. As a result, \baselinename encountered a \texttt{CannotResolveClassException} error, indicating the inability to locate the aforementioned files, thereby failing to discover exploitable vulnerabilities. In contrast, \appname employs the most primitive remote execution method in vulnerability exploitation, directly accessing \texttt{localhost} to ensure the reliable triggering of vulnerabilities. Additionally, in CVE-2020-15250, \baselinename created a test that invoked a private function in JUnit. However, due to the method's inaccessibility in the server project, this test could not evaluate the exploitability of the vulnerabilities.

Moreover, the utilization of migration ensures the stability of vulnerability triggering and proves more effective than the code behavior-guided approach employed by \baselinename. This migration strategy guarantees \appname's performance in complex scenarios. For example, in CVE-2019-12415, reproducing the vulnerability requires passing a specially crafted .xlsx file with a remote execution address included in the file header. Generating such a file through genetic algorithms poses significant challenges, resulting in failures for EvoSuite. However, \appname directly triggers the vulnerability by passing a pre-built exploit file, demonstrating its ability to reliably exploit the vulnerability. In complex scenarios, the migration strategy effectively ensures the reliability of vulnerability triggering.

After analyzing the above conclusions, \appname ensures the stability of reproducing test vulnerabilities through a migration-based strategy. This is particularly crucial when dealing with complex scenarios involving parameters such as files. Additionally, \appname improves the performance of generating tests on the modified call chains by applying rules to modify the passed parameters during the transmission process. Defining the manifestation of vulnerability triggers reduces the need for manual checks and minimizes the impact on the methods.

\subsection{Threats to Validity}

\textbf{External Validity.} One possible factor that may affect the authenticity of the experiment is that the dataset is manually selected, raising concerns about the coverage of vulnerability domains and exploit types. Additionally, We limit our selection to projects with reachable exploit code, which may also be a limitation. Some trigger conditions are not covered in our method, which means a manual check is also required.

\textbf{Internal Validity.} In certain instances, \appname encounters difficulties in generating suitable tests due to the following reasons: 1) Project-related issues, such as the inability to analyze jar files, failures in loading classes in EvoSuite, and uncompilable projects, result in 5 failure cases; 2) the generated tests failed in covering vulnerable function. For example, there is a failure in vulnerability CODEC-263 within the project named java-algorand-sdk. In this case, an if-branch checks one of the input values' types, which should be a valid image format, before reaching the vulnerable function. However, our generated test passes an illegal input, thus preventing the reaching of the vulnerable function.

Despite the aforementioned limitations, our method produces significantly better results than our baseline in a dataset with sufficient types and quantities of vulnerabilities. We are confident that our method has considerable generalizability.
\section{Related Work}

\vspace{0.1cm} \noindent {\bf Software Composition Analysis. }
Software composition analysis is a domain that involves managing vulnerable dependencies in software projects, which includes identifying, tracking, and resolving such vulnerabilities ~\cite{Decan1}. Prior researchers had cross-referenced the dependency versions used in a project against a database of known vulnerable versions to identify potential library vulnerabilities  ~\cite{Whitesource1, Snyk1}. Methods for dependency analysis would generate false positives ~\cite{Elizalde1, Alfadel1, Decan1}, as less than 1\% of packages have a reachable call path to vulnerable code ~\cite{Mir1}. Certain static analysis methods ~\cite{Ponta1, Nielsen1} that rely on generating a static call graph of a software project and identifying potentially vulnerable functions might produce false positives due to discrepancies between the static call graph and the actual run-time behavior of the program ~\cite{Iannone1}. They did not detect whether a vulnerability is reachable, which means whether an attacker can generate an input that passes to the vulnerable function and triggers the dependency vulnerability ~\cite{Darius1, Ponta2}. Dynamic methods involved running the test cases of a software project to generate a call graph and got the control flow. However, these methods are limited by differences between the test cases and the actual code that can trigger dependency vulnerabilities, as well as by limitations in the test coverage and the ability to trigger vulnerabilities under real conditions ~\cite{Kochhar1, Kochhar2}.

Kang et al. noted that the previously mentioned methods didn't consider control flow and if client projects were able to construct an input to trigger the vulnerability ~\cite{Hong1, Ponta1, Hu1}. SIEGE utilized the coverage of the vulnerable function as a search criterion to generate test cases that call the target function ~\cite{Iannone1}, providing evidence that the library vulnerability can be reached. To meet the requirement of domain knowledge, Kang et al. manually selected test cases added to the library after fixing vulnerabilities, which is related to reproducing the vulnerability. Executing these vulnerability-witnessing test cases helped obtain the performance when the vulnerability is triggered. This performance serves as the criterion for generating test cases that satisfy the triggering condition~\cite{Hong1}.

Different from the aforementioned methods, \appname employs the exploit code of the library vulnerability, which collects from open-source forums, to guide test case generation for a client project, which contains more precise domain knowledge. Additionally, the trigger behavior of the vulnerability is reproducing the vulnerability rather than relying on code performance, providing a more persuasive vulnerability exploit test.

\vspace{0.1cm} \noindent {\bf Test Case Generation. }
EvoSuite is a tool that generates test cases with assertions for classes written in Java code by genetic algorithm ~\cite{Von1, Fraser1}. JUnit tests are represented as individuals and fitness scores are optimized using mutation, crossover, and other operators. Kang et al. and Iannone et al. modified the fitness score in EvoSuite to generate tests that can trigger the library vulnerabilities ~\cite{Hong1, Iannone1}. However, these methods lack domain knowledge, despite Kang et al. use vulnerability-witness tests in vulnerable libraries to guide test generation ~\cite{Hong1, Iannone1}.

In addition to search-based methods, deep learning-based methods are widely used in test case generation. VDiscover used static and dynamic features to predict if a test case is likely to trigger a software vulnerability using machine learning techniques ~\cite{Grieco1}. VulDeePecker initiated the study of using deep learning-based vulnerability detection to relieve human experts from the tedious and subjective task of manually defining features, leading to the design 
 and implementation of a deep learning-based vulnerability detection system ~\cite{Li1}.

In our study, we use target function coverage and the similarity between generated test cases and the exploit code to guide test generation and add an exploit code migration step to the test case generated by EvoSuite, which enables us to modify the entry function's parameters and ensures the vulnerability function to receive a value causing vulnerability triggering.

\vspace{0.1cm} \noindent {\bf Vulnerability Exploit Generation. }
Exploit code~\cite{Bao1} is commonly used to detect vulnerabilities and implement defensive measures by exploiting software vulnerabilities during execution, such as taking control of computer systems, causing buffer overflows, or executing unexpected code~\cite{Arce1}. Over the past decade, a significant amount of research has focused on the generation of exploit code for software projects~\cite{Avgerinos1, Xu1, Dixit1}.

AEG generated exploit code using binary information during code execution, but this approach was not universally applicable ~\cite{Avgerinos1}. Xu et al. used symbolic execution to search the target software and found potential buffer overflow vulnerabilities, then generated the exploit of software vulnerability ~\cite{Xu1}. However, it did not perform well in complicated programs. AngErza used dynamic and symbolic execution to identify hot spots in the code, formulate constraints and generate a payload based on those constraints ~\cite{Dixit1}.

Our work relies on exploit code for the vulnerable library functions to guide test generation. We manually select exploit code from open-source websites due to the diversity of vulnerabilities and the requirement for a clear exploit code.
\section{Conclusion and Future Work}
In this paper, we propose a method \appname which utilizes the genetic algorithm
to generate test cases covering vulnerable functions in the project. By migrating the vulnerability exploit code into the generated test, we construct an exploit test case for the library vulnerability in the project. Compared to \baselinename, our unique migration step results in generating 53.4\% more exploits. Executing the test case provides a reference for the developer to check if the project has exploitable library vulnerability and decide whether to update the dependency version. In the experimental evaluation, we test our method in 60 vulnerability-project pairs and receive 43 exploits, which shows its efficacy.

In the future, we plan to perform more experiments with more datasets to further evaluate the performance of our method. We will also explore the use of Large-scale Language Models to find library vulnerabilities.

\section*{acknowledgement}
This research was supported by the National Natural Science Foundation of China (No. 62141222) and the National Research Foundation, Singapore under its Industry Alignment Fund – Prepositioning (IAF-PP) Funding Initiative. Any opinions, findings and conclusions or recommendations expressed in this material are those of the author(s) and do not reflect the views of National Research Foundation, Singapore.

\bibliographystyle{ACM-Reference-Format}
\bibliography{main}

\end{document}